\documentclass[10pt,conference]{IEEEtran}
\IEEEoverridecommandlockouts
\usepackage{cite}
\usepackage{amsmath,amssymb,amsfonts}
\usepackage{algorithm}
\usepackage{algpseudocode}
\usepackage{graphicx}
\usepackage{longtable}
\usepackage{authblk}
\usepackage{multirow}
\usepackage{mdframed}
\usepackage{textcomp}
\usepackage{booktabs}
\usepackage{svg}
\usepackage{makecell}
\usepackage{adjustbox}
\usepackage[numbers]{natbib}

\usepackage{soul}
\usepackage{listings}
\definecolor{lightgray}{rgb}{0.9, 0.9, 0.9} 
\lstdefinestyle{myStyle}{
    backgroundcolor=\color{lightgray}, 
    basicstyle=\ttfamily\scriptsize, 
    frame=single, 
}
\usepackage{xcolor}
\def\BibTeX{{\rm B\kern-.05em{\sc i\kern-.025em b}\kern-.08em
    T\kern-.1667em\lower.7ex\hbox{E}\kern-.125emX}}
\begin{document}

\title{Unified Semantic Log Parsing based on Causal Graph Construction for Attack Attribution}
\author{Zhuoran Tan\thanks{Email: z.tan.1@research.gla.ac.uk}}
\author{Christos Anagnostopoulos\thanks{Email: Christos.Anagnostopoulos@glasgow.ac.uk}}
\author{Shameem P. Parambath\thanks{Email: Sham.Puthiya@glasgow.ac.uk}}
\author{Jeremy Singer\thanks{Email: jeremy.singer@glasgow.ac.uk}}
\affil{School of Computing Science, University of Glasgow, UK}
\affil{\textit{\{z.tan.1\}@research.gla.ac.uk, \{christos.anagnostopoulos, sham.puthiya, jeremy.singer\}@glasgow.ac.uk,}}

\maketitle

\begin{abstract}
Multi-source logs provide a comprehensive overview of ongoing system activities, allowing for in-depth analysis to detect potential threats.
A practical approach for threat detection involves explicit extraction of entity triples (subject, action, object) towards building provenance graphs to facilitate the analysis of system behavior. However, current log parsing methods mainly focus on retrieving parameters and events from raw logs while approaches based on entity extraction are limited to processing a single type of log.
To address these gaps, we contribute with a novel unified framework, coined UTLParser. 
UTLParser adopts semantic analysis to construct causal graphs by merging multiple sub-graphs from individual log sources in labeled log dataset. It leverages domain knowledge in threat hunting such as Points of Interest. We further explore log generation delays and provide interfaces for optimized temporal graph querying. Our experiments showcase that  UTLParser overcomes drawbacks of other log parsing methods.
Furthermore, UTLParser precisely extracts explicit causal threat information while being compatible with enormous downstream tasks.
\end{abstract}


\section{Introduction}
Effective and comprehensive representations of information within raw data is crucial for timely and accurate threat detection. 
Recent studies have focused on extracting event templates and parameters from logs \cite{9134790, 10109145}.
These approaches show commendable performance in terms of processing time and precise extraction of event templates. They mainly extract parameters aimed at anomaly detection. However, as parameters alone lack semantic context, direct anomaly detection becomes ineffective.
A few studies have explored using temporal graphs for log representation to detect changes in system status. 
For example, the graph-based methods in \cite{kingEulerDetectingNetwork2023,cai_structural_2020} detected abnormal changes in temporal network status. However, such approaches rely strictly on exact timestamp matching and overlook the fusion of multi-source data and spatio-temporal anomaly properties. 


Since advanced anomalous behavior involves multiple indicators, an effective approach must uncover causal relationships among Indicators of Compromise (IOCs) across multiple logs.
Assessing the temporal properties of causal graphs is critical for identifying current system status and enabling rapid detection.
In this paper, we address the aforementioned challenges, which can be grouped as follows:
\begin{itemize}
\item{} \textbf{Dependency on Single Data Source:~} Precise detection of abnormal system statuses requires information from multiple data sources.

\item{} \textbf{Semantic Incoherence:~} Unlike semantically coherent and contextually meaningful sentences, logs typically consist of short, fragmented and non-coherent entries. 
\end{itemize}

\textbf{Contribution:} Our generic log processing framework adapts to various log semantic structures leveraging domain knowledge to extract essential components. Those components are defined as Points of Interest (POIs) referring to fields or locations where IOCs can be detected. 
Our contribution is an extensible and scalable \textbf{U}nified \textbf{T}emporal \textbf{L}og \textbf{P}arser (\textbf{UTLParser}) for multi-source log fusion. UTLParser employs a fused, semantic dependency-based approach to construct comprehensive causal graphs from labeled log datasets. 
\textbf{Note:} The source code is publicly available \cite{tan_2024_14197366} for reproducibility, including the benchmark testing, and dataset.
\section{Related Work}

Semantic log parsing has been extensively explored in the cyber-security domain for threat intelligence analysis. 
Several approaches focus on extracting actionable entities from audit logs
\cite{milajerdi_holmes_2019} 
or process execution logs, 
\cite{10.1007/s44227-023-00014-9}
and then constructing causal graphs. 
In general, these methods can be broadly classified 
as either based on \emph{(i)} static rules or \emph{(ii)} models learned from data.
Deep Neural Networks (DNNs) have been widely used in semantic analysis. 
Semparser \cite{huo_semparser_2023} implemented a DNN to predict the semantic roles of tokens in logs. EdgeTorrent \cite{king_edgetorrent_2023} leveraged a DNN for graph embedding over a stream of edges derived from `syslog' to construct dynamic networks of entity interactions. 
CyberEntRel \cite{ahmed_cyberentrel_2024} utilized the attention-based RoBERTa-BiGRU-CRF model to tag sequences and extract relation triples using relation-matching techniques.
Rule-based approaches have been used to find the dependency between entities. 
HOLEMS \cite{milajerdi_holmes_2019} outlined correlations between suspicious information flows and mapped low-level entities from audit logs to High-Level Scenario Graphs (HSG) to detect Advanced Persistent Threats (APTs).
POIROT \cite{milajerdi_poirot_2019} proposed a causal correlation-aided semantic analysis method to automatically mine the causality among anomalous events for threat hunting. These works have achieved high prediction performance, but are limited by the generality to other log types. 
Note,  
the testing datasets in all approaches are relatively narrowed down to a single type of data source. This limitation potentially leads to the omission of other critical information that could help identify stealthy attacks. 
In UTLParser, the construction of causal graphs results in directed multi-edge graphs.
It can be utilized in various downstream anomaly detection methods, including pattern-based, scoring-based, or even temporal snapshot-based anomaly detection. 
We encompass log data, domain knowledge, network traffic, and process execution information without focusing on specific attacks or single data sources, aiming to comprehensively demonstrate system's operational status.

\section{Methodology}
\begin{figure*}[ht]
    \centering
    \includegraphics[scale=0.6]{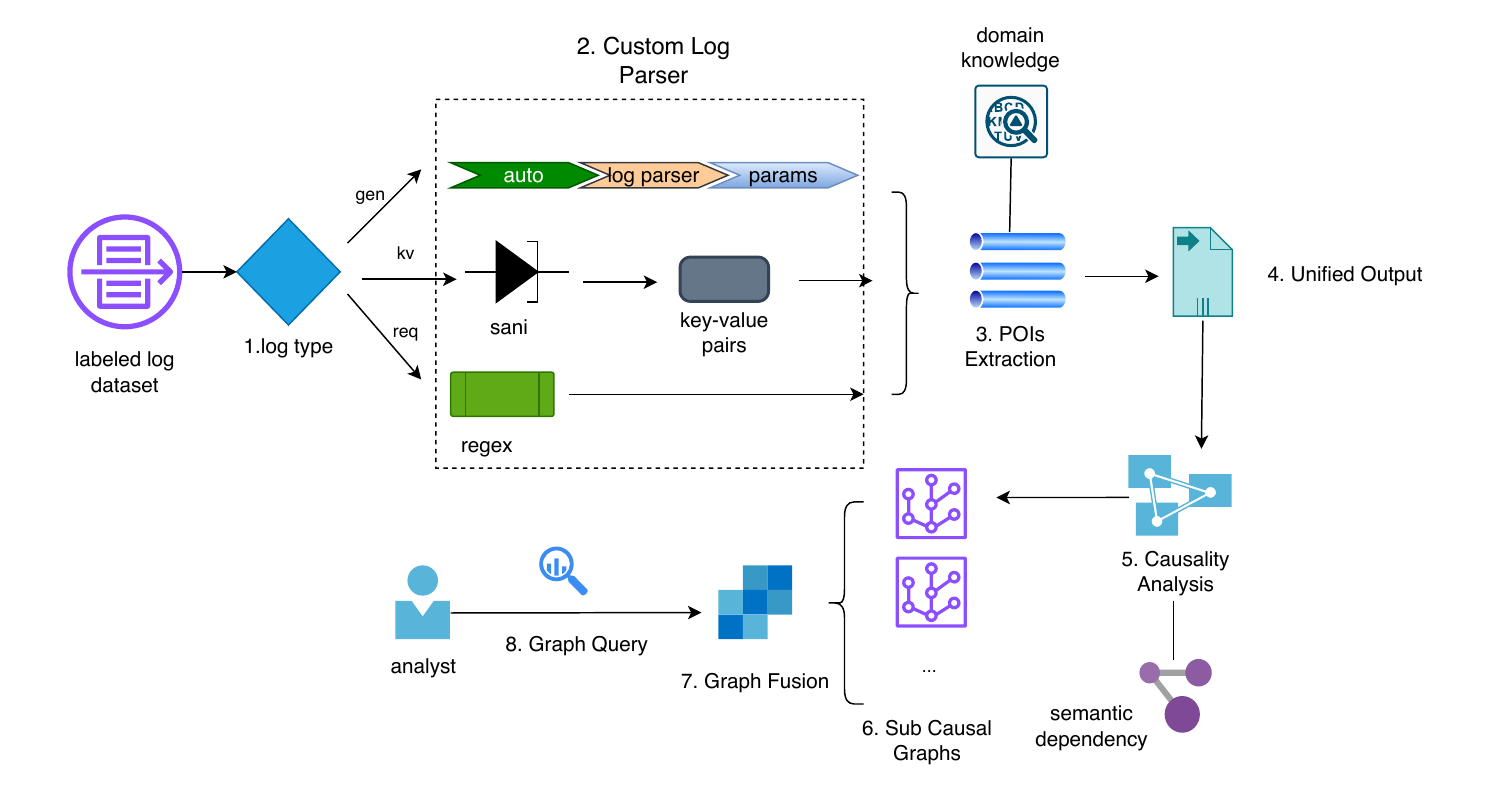}
    \caption{Unified Log Parsing Framework (UTLParser). Steps: 1. Check log type; 2. Parsing logs; 3. Extraction of POIs; 4. Output with unified format; 5. Causality analysis upon semantic dependency; 6. Causal graph construction; 7. Fuse sub causal graphs; 8. Graph query via optimized timestamp.}
    \label{fig:causal graph framework}
\end{figure*}

UTLParser constructs causal graphs from labeled log datasets using sequential pipelines, as shown in Fig. \ref{fig:causal graph framework}. 
It provides a multi-source, log-based parsing solution to handle the diverse log types existing in the target dataset.
Prior to log parsing, UTLParser verifies the log type to trigger specific parsing logic. 
Before generating unified output, UTLParser leverages predefined domain knowledge to extract the desired information contributing to threat hunting and activity representation.
By performing a semantic analysis of the output data, the system can identify actionable verbs and noun tokens as entities. 
Furthermore, we use SemgrexPattern \cite{10.1145/2063576.2063763} to determine the dependency relationship between entities
and generate causal sub-graphs as output. 
Our system consolidates all the individually generated sub-graphs into a comprehensive graph for downstream analysis tasks.

\subsection{Dataset Description}
We use two datasets to cover both semi-structured and structured log parsing for constructing causal graphs. The Austrian Institute of Technology (AIT) dataset \cite{landauer_2022_5789064} is a semi-structured log dataset, which we process to generate fused causal graphs. 
The Internet of Things (IoT)-32 dataset \cite{sebastian_garcia_2021_4743746} consists of network traffic-based data in a structured log format.
Both datasets are labeled and contain multiple instances of attacks.
To create comprehensive IOCs and POIs, we refer to the ground-truth labelling methods introduced in original dataset. We then extract part of logs used for labelling anomalies in \cite{landauer_2022_5789064} to evaluate UTLParser and benchmark. These extracted logs demonstrate extinct patterns and are suitable for extracting entities from them. In addition, we also collect various large-scale logs from \cite{landauer_2022_5789064} to perform scalability tests. 



\subsection{Log Parsing}
Log parsing is the core component of our pipeline. 
The goal is to extract meaningful tokens that represent entities such as subjects, actions, and objects. 
To invoke the correct log processing logic, 
we pre-define a configuration file specifying log types to their categories. UTLParser decides the log category when reading log filenames.
The logs are grouped into three main categories: \emph{(i)} general logs, \emph{(ii)} key-value logs and \emph{(iii)} request logs. 

\paragraph{General Log Parser}
General logs are defined according to the common patterns that exist in the logs. Typically, they include information like timestamp, protocol, and content. We refer to implementation in \cite{He2017DrainAO}, which parses logs with a tree structure based on a similarity threshold and fixed depth, and extends it with automatic parameter selections.

\paragraph{Key-Value Pair Log Parser}
Key-value logs follow a key-value structure, either partially or fully. They are processed in a tokenized manner and then parsed through key-value patterns. 
The outcome is dictionaries of key-value pairs. 

\paragraph{Request Log Parser}
Request logs are the web-service-facing records.
The access log is one typical example of this type. 
In such logs, it is more efficient to identify malicious behavior by extracting the parameters directly rather than mining for event templates. UTLParser applies regular expressions (regex) to match the parameter parts of URLs and specific user agent names.

\subsection{Points of Interest}
POIs denote potential locations where existing IOCs might be found. We define them as references to extract critical parts of logs.
In this framework, all IOCs are extracted from pre-defined labeling rules \cite{landauer_2022_5789064, sebastian_garcia_2021_4743746}, corresponding to specific IPs, event types, and ports, etc. 
Regarding POIs, we have formalized all necessary information from logs into unified column names, which are as follows:
\begin{lstlisting}[style=myStyle]
"Time", "Src_IP", "Dest_IP", "Proto", "Domain", "Parameters", "IOCs", "PID", "Actions","Status", "Direction"
\end{lstlisting}
We define them as the places to find potential IOCs. 

\subsection{Dependency Extraction}
After parsing the logs, we consolidate all essential and relevant information into a unified output format. Semantic analysis is then applied to this output to extract dependencies. This process relies on the assumption that specific semantic properties exist within entity triples in certain logs. Typically, the action corresponds to a verb token, while the subject and object are represented by noun tokens. These tokens align with predefined IOCs, and the extraction points for these tokens are identified as POIs.

Semantic tags and dependencies are extracted using SpaCy\footnote{https://spacy.io/}, with dependency parsing identifying subgraphs comprising an action (parent node) and its subject/object (child nodes). Relationships between tokens are inferred using SemgrexPattern, and the final causal graph integrates these subgraphs with predefined node and edge attributes.

\subsection{Graph Fusion}
Graph fusion synthesizes knowledge from sub-graphs after the initial causal graph construction. Due to the fact that one event can trigger the generation of multiple log types, fusing diverse log sources provides an efficient way to analyze potential malicious behavior.
Consider a series of sub-graphs $ \{G_i | i = 0, 1, \ldots, n\}$, where each sub-graph $G_i = (V_i, E_i) $ comprises its own set of nodes ($V_i$) and edges ($E_i$). The fused graph is denoted as $R = (V_R, E_R)$, where
$V_R$ consists of $V_R = \bigcup_{i=0}^{n} V_i$ while $E_R = \bigcup_{i=0}^{n} E_i$. 
The attributes of edges and nodes are preserved during the fusion process. 
However, if a node or edge appears in multiple sub-graphs, the attributes in the final fused graph will be those from the last subgraph in which that node or edge appears. 
The update process is:
$\text{attr}_V(v) = \text{attr}_{V_i}(v) \text{, for } i = \max \{ j \mid v \in V_j \}$, and
$\text{attr}_E(e) = \text{attr}_{E_i}(e) \text{, for } i = \max \{ j \mid e \in E_j \}$.

\subsection{Graph Query}

A graph query provides an interface for generating temporal graphs from a fused graph. Although temporal graphs can be extracted using the exact timestamps that match, there is a delay for events to log in different dimensions due to system settings, custom logging frequency, and network status \cite{10.1016/j.is.2023.102246,landauer_have_2021}.
In order to correlate all related recordings, 
UTLParser introduces a delay tolerance $\Delta t$, to contain as much information as possible without breaking the independence of individual events. 
The queried event is thus within a time period range $\{T - \Delta t, T + \Delta t\}$ instead of the exact timestamp $T$.
The delay tolerance considers two factors: graph integrity and graph independence. 
Graph integrity indicates whether the extracted graph has included complete recordings of one event, which determines the integrity of the event trace. 
Graph independence measures the noise existing inside the graphs, ensuring the independence of extracted information only relating to specific events. The detailed algorithm to calculate two factors is represented in Algorithm \ref{alg:temp_graph}.

\begin{algorithm}
\caption{Temporal Graph Extraction}
\label{alg:temp_graph}
\begin{scriptsize}
\begin{algorithmic}[1]
\State \textbf{Input:} Sub Graph List $G_L$, Timestamp $T$, Time Delay Candidates $t\_l$, Average Length $\overline{L}$
\State \textbf{Output:} Optimized Temporal Graph $G_{T \pm \hat{t}}$, $\hat{t}$ is calculated optimized time delay
\State // Step1: Connect all sub graphs into one graph
\State $G \gets \text{Graph\_Compose($G_L$)}$   

\State // Step2: Choose optimized time threshold
\State $score\_dict \gets \{\}$
\For{$t \text{ in } t\_l$}
    \State $G_{T\_t} \gets temp\_graph(G, T, t)$
    \State $score \gets inte\_score(G, \overline{L}) + inde\_score(G)$
    \State $score\_dict \gets t, score$
\EndFor
\State $\hat{t} \gets min(max($score\_dict$, key=value), key=key)$

\State // Step3: Extract optimized temporal graph
\State $T_b, T_u \gets \text{Time\_Scope(T, t)}$ \Comment{calculate the time scope}
\For{$u, v, edge \text{ in } G.edges$}
    \If{$edge[T] \geq T_b \text{ and } edge[T] \leq T_u$}
    \State $G_{T \pm \hat{t}} \gets \text{add\_edge}(u,v, **edge)$
    \EndIf
\EndFor
\State \Return $G_{T \pm \hat{t}}$
\end{algorithmic}
\end{scriptsize}
\end{algorithm}

\begin{table*}[htbp]
\caption{Parsing Accuracy and F1 Scores for Log Parsing Across Four Data Types}
\label{tab:benchmark}
\centering
\begin{tabular}{lcccccccccccc}
\hline
\multicolumn{13}{c}{\textbf{Parsing Accuracy and F1 Scores}} \\ \hline
\multirow{2}{*}{\textbf{Log Parser}} & \multicolumn{4}{c}{\textbf{Parsing Accuracy}} & & \multicolumn{4}{c}{\textbf{Parsing F1 Score}} & & \multicolumn{2}{c}{\textbf{Average}} \\ \cline{2-5} \cline{7-10} \cline{12-13}
 & \textbf{auth} & \textbf{audit} & \textbf{dns} & \textbf{syslog} & & \textbf{auth} & \textbf{audit} & \textbf{dns} & \textbf{syslog} & & \textbf{Accuracy} & \textbf{F1 Score} \\ \hline
Logram \cite{9134790}   & 0.0053 & 0.0060 & 0.0130 & 0.2775 & & 0.4052 & 0.4052 & 0.4052 & 0.4052 & & 0.0755 & 0.4052 \\ \hline
NuLog \cite{nedelkoski2020selfsupervised}    & 0.0027 & 0.1385 & 0.0270 & 0.6555 & & 0.4866 & 0.4866 & 0.4866 & 0.4866 & & 0.2059 & 0.4866 \\ \hline
ULP \cite{9978179}     & 0.0053 & 0.3425 & 0.0270 & 0.6593 & & 0.9750 & 0.9750 & 0.9750 & 0.9750 & & 0.2585 & 0.9750 \\ \hline
Brain \cite{10109145}   & 0.9973 & 0.3410 & 0.9880 & 0.3971 & & 0.6538 & 0.6538 & 0.6538 & 0.6538 & & 0.6808 & 0.6538 \\ \hline
UTLParser & \textbf{0.998} & \textbf{0.999} & \textbf{0.988} & \textbf{0.9455} & & \textbf{0.998} & \textbf{0.999} & \textbf{0.9999} & \textbf{0.9979} & & \textbf{0.9826} & \textbf{0.9984} \\ \hline
\end{tabular}
\end{table*}

\section{UTLParser Tool Technical Description}

Detailed running instructions and environment configuration have been included in \cite{tan_2024_14197366}. It can be downloaded using 
\texttt{git clone https://github.com/Wapiti08/UTLParser.git}. Pyenv \footnote{https://github.com/pyenv/pyenv} is used for configuring virtual environment. All experiments are performed with Python version 3.10 on macOS with M2 chip and 32G memory. After activating virtual environment, specific dependent libraries have been included inside \texttt{requirements.txt}.  

The running entry-point locates at src folder with the filename of \texttt{main.py}. We define multiple parameters inside for diverse functional running. The core parameters include:

\textit{- a (application)} indicates the type of log to process.

\textit{- e (entities)} specifies the list of desired entity types to extract, choose from config regex keys. The default values are "ip4" and "domain".

\textit{- al (app\_list)} specifies the list of application names (log types) to process. It is often used together with another parameter \textit{- f (fuse)} with \textit{True} option.

\textit{- t (timestamp)} specifies the timestamp to query the temporal graph, using with \textit{- al} to indicate the sub-logs of fused graph.


\section{Performance Evaluation}
To evaluate log parsing performance, we create sample auth, audit, syslog, and dns logs from \cite{landauer_2022_5789064}. The process involves hand-crafted correction after initial parsing trail.

\subsection{Log Parsing}

We conducted the log parsing experiment with parsing accuracy and parsing F1 score on benchmark datasets, following the same evaluation methods introduced in \cite{10.1109/ICSE-SEIP.2019.00021}. 
Let $T_p$ represent the number of true positive pairs (accurate pairs), $N_p$ be the total number of pairs parsed, $N_r$ the total number of real pairs, and $T_e$ the number of accurate events, while $N_s$ represents the total number of events in the ground truth series.
The parsing accuracy is defined as $\frac{T_e}{N_s}$, 
and parsing F1-Score is denoted as $ F_1 = 2 \cdot \frac{\text{Precision} \cdot \text{Recall}}{\text{Precision} + \text{Recall}}$
Notably, these metrics are different from general metrics in machine learning application measurement.
We prepare 2000 audit, 748 auth, 2000 DNS, 1045 syslog, to evaluate log parsing. 


As shown in Table \ref{tab:benchmark}, we compare the performance of audit log parsing across four art algorithms. UTLParser outperforms all other methods across all datasets and exceeds the baseline performance in every metric. In particular, on the auth logs, UTLParser resulted in near 100\% detection. 
This is attributed to the simple structure of auth logs, where variables such as `username' and UID within the log content can be identified as parameters via tree parsing. Performance in audit logs relies on exact regex-based matching of keys and values. In addition, other approaches (DNNs based, tree parsing, etc) are constrained by the testing size of logs and unstable threshold choices when applied to key-value pairs based logs.

\subsection{Scalability}

\begin{table}[htbp]
\centering
\caption{Evaluation of Processing Time (s) at Stages}
\label{tab: processing time evaluation}
\begin{tabular}{ccccc}
\hline
\textbf{\thead{Dataset}} & \textbf{Size} & \textbf{\thead{Semantic Log \\ Parsing}} & \textbf{\thead{Causal\\ Graph}} & \textbf{\thead{Graph \\Fusion}}  \\ \hline
access                    &   412922        &   8.5        &  72.9    & \multirow{3}{*}{180.9}             \\ \cline{1-4}  
audit                     &   330990         &   14.1   &     53.6                  &                    \\ \cline{1-4}
dns                   &   335875          &   849.5      &   63.2                    &                                  \\ \hline
\end{tabular}
\end{table}

To evaluate scalability, we collected logs in three main categories: general logs (dns), key-value logs (audit) and request logs (access). 
As shown in Table \ref{tab: processing time evaluation}, log parsing for dns took the longest time at 849.5 seconds, compared to shorter times for the other logs. This is because dependency analysis is applied individually to the dns logs, whereas default dependency relations are used for access and audit logs. The time required for the construction of the causal graph was approximately one minute for all logs. The final graph fusion process took only three minutes, which is acceptable for processing large volumes of data on a daily basis.

\section{Limitations}
The main function of the proposed tool is to convert a labeled log dataset into temporal causal graphs. However, due to the initial design, some manual configuration of IOCs (Indicators of Compromise) and POIs (Points of Interest) is still necessary. To enhance user experience, one of the future plans is to streamline these configurations using automation technologies. Another limitation arises from the chosen dataset, as the log types may cover only a limited range. While the logs are representative, other categories might be missing. To address this, we plan to expand the log pool through further use of UTLParser and feedback from potential users.


\section{Conclusions}
We investigate semantic entity triple extraction from multi-source logs. We propose UTLParser, a framework to semantically parse logs to causal graphs and achieve the fusion of subgraphs to one comprehensive graph. This graph is a 
directed multigraph
capturing all timestamps and potential POIs as node or edge value and attributes. We implement interfaces
for 
temporal graph extraction and corresponding sub-graph level labeling. Through extensive experiments, we prove the precise parsing performance compared with other 
approaches and demonstrate a low missing rate of critical IOCs information after transformation.
Our research agenda includes exploration regarding graph compression and further reduction of the graph size. 
This work can be leveraged for direct graph-based anomaly detection and temporal graph learning.

\section*{Acknowledgment}
This research has received funding from JUMPSEC Ltd.
\clearpage





\end{document}